# Quasi-static perforation of the ceramic foam core sandwich panels impregnated with Newtonian and non-Newtonian fluids

Ali Afrouzian[a*], Salman Pouryazdian[b], GolamHossein Liaghat[b,d], AhmadReza Bahramian[c]

a W. M. Keck Biomedical Materials Research Lab, School of Mechanical and Materials Engineering, Washington State University, Pullman, WA, 99164, USA

b School of Mechanical Engineering, Tarbiat Modares University, Tehran, Iran

b Polymer Engineering Department, Faculty of Chemical Engineering, Tarbiat Modares University, Tehran, Iran

d School of Mechanical and Automotive Engineering, Kingston University London, London, United Kingdom

**Abstract:**

In this research, Mechanical response and energy absorption of ceramic foam sandwich panels filled with Newtonian and non-Newtonian fluid subjected to quasi-static indentation loads were investigated experimentally. In addition, the effects of number of face-sheet layers (face-sheet thickness) and boundary conditions on the mechanical response and energy absorption of sandwich panels were discussed. Sandwich panels consisted of two glass/epoxy face-sheets with different thicknesses (4 and 8 layers) and an open cell ceramic foam core. This foam saturated with Newtonian (Glycerol) and non-Newtonian (mixture of Ethylene glycol and Aerosil) fluids. Sandwich panels were tested in two different kinds of supporting conditions, simply-supported and fully fixed (clamped). Results showed that damage area in neat composite and STF filled sandwich panels are much higher than the glycerol filled foam core sandwich panels. Also, STF saturated sandwich panels absorbed higher energy compared with two other type of sandwich panels during quasi-static punch shear test.

[*] Corresponding Author: ali.afrouzian@modares.ac.ir

**Keywords**: Indentation shear test; Composites; Ceramic foam; Rheology; Energy Absorption

# 1. Introduction

Sandwich structures consist of two stiff and strong face sheets separated by a lightweight core, which has lower strength compare to the skins. These structures have been widely used in different applications such as aerospace, portable structures and sport equipment [1-3]. The separation of the two faces by the core (metal, foam, and honeycomb) intensifies the second moment of area of the panel with little increase in weight, producing an efficient structure for tolerating bending and buckling loads [4]. One of the main limitations of sandwich structures and composites is their vulnerability to the transverse loading such as impact and quasi-static perforation [5-13].

Erkendirci et al. [14] studied the quasi-static perforation of PW S-2 Glass/SC-15, PW S-2 Glass/HDPE and PW E-Glass/HDPE composites and considered them at varying thicknesses. The maximum load carrying capacities were figured out to change with the support span radius (SPRs), and for the HDPE composites were lower than SC-15. Also, they presented that HDPE systems have lower pick force, lower stiffness, higher deflection, and lower damage area compared to SC-15 composites.

The use of quasi-static penetration tests to predict impact behavior of the composite materials has been studied [15-19]. Sutherland and Soares [17] evaluated the initial impact behavior and onset of the delamination through the static tests to predict low velocity impact behavior of the marine GRP laminates. They identified that in both dynamic and static tests, damage progression was initiated with a sudden appearance of internal delamination. This kind of damage was followed by fiber fracture when the incident energy reached to its highest value.

Feli et al. [20] modified and analytical model for clamped circular sandwich panels subjected to the low velocity impact. Maximum contact force, deflection of the central point, energy absoption and contact duration were analyzed. In addition, their model has capability to predict low velocity impact response of other sandwich panels with different cores such as foam and truss.

The comparison between energy absorption of skin laminated and foam back laminated in a composite structure when they were subjected to a static and impact loadings have been presented by Roach et al.

[21]. They concluded that skin laminate absorbed considerable amount of energy during penetration by deflection, which this behavior was not observed in the foam core. They also found that designing a composite laminate with several thin laminates separated with air space is more convenient and absorb more energy during penetration rather than single thick laminate.

Effect of the core properties including its thickness, material properties, and geometry on the energy absorption and damage mechanisms of the sandwich structures brought large number of researches [22-26].

Hassan et al. [23] investigated variation of the properties of polymeric foam on the perforation resistance of a glass fiber/epoxy sandwich structures by measuring the principle mechanical properties of various foams and their effects on the perforation resistance. Mode I and mode II tests have been carried out in order to obtain fracture behavior of these structures. They eventually found that mode II fracture behavior is much greater that (up to thirty times) mode I, proposing this may be a prominent energy absorbing failure mode in perforated sandwich panels.

Energy absorption and specific energy absorption of the foam core sandwich panels subjected to the quasi-static indentation have been carried out by Li and his collaborators [27]. They considered face-sheet and core thickness, their relative densities and nose shape of the indenter for their research. They concluded that cores higher density could increase energy absorption as well as face thicknesses.

Jing et al. [28, 29] studied the mechanical behavior of the three different foam core sandwich structure subjected to impulsive loading. Closed-cell aluminum foam core, open-cell aluminum foam core, and aluminum honeycomb core have been used. They found the basic failure mechanisms of the closed-cell foam core sandwich panels which were global inelastic deformation in addition with compressive failure beyond the projectile. In addition, in the case of energy absorption, honeycomb core was the best

One of the main failure modes in the foam core sandwich structures during impact test is shear failure [30]. They simulated composite structures behavior using ABAQUS/Explicit software and understood there is a precise agreement between numerical and experimental investigations. Moreover, they figured out that the linear cross-linked PVC foams presented higher perforation resistance compared with linear PVC foams. It has also represented that placing a high density foam core below the top surface skin can improve perforation resistance compared with sandwich panels in which the higher density foam used in the lower skin surface.

Low velocity impact on the foam core sandwich panels with different densities and using carbon fibers in their surface have been carried out by Zhou and his collaborators [26]. Force-displacement curve and failure modes have been summarized experimentally and numerically. They documented that shear failure was the main failure mode in the sandwich composite structures.

Finite element analysis has been used to simulate damage response of the foam core sandwich structures subjected to low velocity impact [31]. Some of the failure mechanisms such as fiber fracture, matrix cracking and delamination were modeled and Crushable foam plasticity, modeled the nonlinearity of the foam's behavior.

Ruan et al. [32] studied the mechanical response and energy absorption of aluminum foam sandwich panels subjected to quasi-static indentation. They considered the effect of different parameters such as the type of face-sheet and their thicknesses, boundary conditions, and surface condition of the face-sheets on the energy absorption and mechanical properties of aluminum foam sandwich panels in their investigations.

In Schubel et. al [33] low velocity impact behavior of sandwich panels consisting of woven carbon/epoxy face-sheets and PVC foam core investigated experimentally. They identified that equivalent quasi-static tests can predict low velocity impact behavior of the structure except for localized damage.

Yu et al. [18] used quasi-static and low velocity impact tests in the sandwich beams with aluminum-foam core to investigate their deformation and failure behavior. They discovered that the failure mode predicted by Gibson's theory agree well with quasi-static test data. They also found that when the impact velocity is lower than 5 m/s, the quasi-static model can forecast the dynamic failure mode at the early stage.

Shear thickening is a non-Newtonian behavior described as the increase of viscosity with the enhancement in the applied shear rate [34]. Some researchers used this behavior as a key element, to boost perforation resistance of composite structures, especially when they were subjected to transverse loadings [35-39].

Hassan et al. [34] dispersed nano-silica particles in the polyethylene glycols using ultrasound mixer. Kevlar and Nylon fabrics were soaked in STF/ethanol solution to construct STF composites and then quasi-static penetration test were conducted on them. They discovered that STF composites have greater resistance to penetration rather than neat composites. In addition, they found that STF saturated fabrics and neat composite have the same flexibility.

Puncture resistance performance and frictional behavior of the high modulus polypropylene fabrics saturated with STF, containing carbon nanotubes were studied [40]. Results showed that STF-treated fabrics increase the puncture resistance of the composite. With the presence of CNTs in the suspension, puncture resistance has showed lower enhancement due to the level of shear thickening which has been decreased.

Fahool et al. [41] carried out friction, quasi-static, and high velocity impact test to study the mechanical behavior of Twaron fabric saturated with shear thickening fluid (STF). They used silica oxide Aerosil OX50 in 35, 40, and 45 wt% as a suspending materials dispersed in polyethylene glycol. By conducting SEM and TEM studies, they established good dispersion of nano-silica particles in the suspension has obtained. They also found from drop weight impact test, open cell foams impregnated with STF have good energy absorption capability. On the contrary, in the high velocity impact test, STF saturated ballistic fabrics, represented same level of performance as dry fabric based on its weight.

Numerical investigation of ballistic behavior of STF impregnated Kevlar fabrics have been proposed using commercial software tool LS-DYNA with the modeling established based on friction property values acquired experimentally in literature for STF impregnated fabrics [42]. They assumed friction as the main energy absorption mechanism in their simulation, but they found contrasting energy absorption trends compared with previous experimental research. As a result, they figured out other prominent energy absorption mechanisms should be taken into consideration in the numerical studies.

Lu et al. [43] studied impact behavior of warp-knitted spacer fabrics (WKSFs) impregnated with shear thickening fluid subjected to low velocity impact loadings numerically using ABAQUS/Explicit software. There model was based on microstructure and fibers were modeled as tic-plastic solid, while STF supposed as power-law fluids. They investigated that the crucial factor that affect the composites was the coupling effect between the STF and fiber tows which contributes to higher stiffness and lower pick force in the STF saturated fabrics compared with the neat ones.

Some other researchers investigated the rate dependent behavior of foam based composite structures saturated with Newtonian and non-Newtonian (STF) fluids [[44-46]]. They also studied the comparison of energy absorption between dry foam, Newtonian-filled foam, and non-Newtonian filled foam and discovered the third group of composite structures, have demonstrated an increase in energy absorption capabilities.

The objective of this study is to develop a novel foam core sandwich panels impregnated with shear thickening fluids. Quasi-static indentation tests have been conducted on the foam-based sandwich structures filled with Newtonian and non-Newtonian fluid in order to investigate puncture resistance of these structures. Also, the effect of composite layer thickness of the face-sheets, boundary condition of the fixtures, and STF saturated fabrics on the energy absorption, damaged areas, and deformation of the skins and core have been analyzed experimentally.

## 2. Experimental Procedure

### 2.1. Materials

Sandwich structures were made of woven E-glass fiber with surface density of 200 g/m2 and epoxy resin consist of diglycidyl ether of bisphenol A (Epon 828) resin and TETA as the curing agent supplied by Shell with a mixing ratio of 1:10, respectively. Core material was ceramic foam (300 kg/m$^3$) has 25 mm thickness and composed from Alumina and Silica fiber manufactured by Luyang. The STF suspension was made from Ethylene glycol (EG) with density 1.113 g/cm$^3$ obtained from Dr. Mojallali company (Iran) as a basic fluid, and fumed silica powder by the name of Aerosil 200 from Degussa. Aerosil 200 has the specific surface area of 200 m$^2$/g with average size of 12 nm, and density of 50 g/l. Glycerol with the specific density of 1.26 g/cm$^3$ was used as Newtonian fluid in this research.

## 2.2. Synthesis of STF

The STF was fabricated using combination of mechanical and ultrasonic mixing. The initial mixing was carried out using a vortex mixer with the speed of 1100-1300 rpm. In this stage, small amount of fumed silica (about 5% of the total weight of Aerosil) were added to the EG at separate stages and mixed together around 20-80 min until a visually homogeneous suspension was acquired. At the end of the first stage of mixing, another mixing have been conducted using high speed Shear mixer with the average speed of about 10000-15000 rpm to reduce the viscosity of the suspension and make sure they reached a homogenous mixture. Ultrasound sonication (with the power of 150 W/cm2 and 24 kHz frequency) for 20 min helped to increase the distribution of the particles within the suspension and prevented nanoparticles from agglomeration. The final mixture was outgassed using vacuum oven for 15 min at 60°C to bring out air bubbles from suspension and aggregation [47]. STF suspensions were prepared at three different concentration of nano-silica particles: 20%, 27.5%, and 35% wt.%, respectively. Four stages of STF preparation are presented in Fig. 1.

## 2.3. Rheological test

Rheological test were executed on STF and EG samples to investigate the effect of shear rate on viscosity. The tests were conducted using a rotational rheometer (MCR300 manufactured by Anton paar) at room temperature (25°C). The plate diameter was 35mm with the angle of 2.5°. The shear rate range was 0-1000 s-1 and the corresponding viscosities were recorded.

## 2.4. Composite fabrication

Hand lay-up procedure was used to manufacture composite plates with 4 and 8 layers of 2D woven glass fibers. The composites layers cured at room temperature (25°c) for one week at the constant pressure to reach a high quality surface, and then were cut in a circular shape in the diameter of 12 cm.

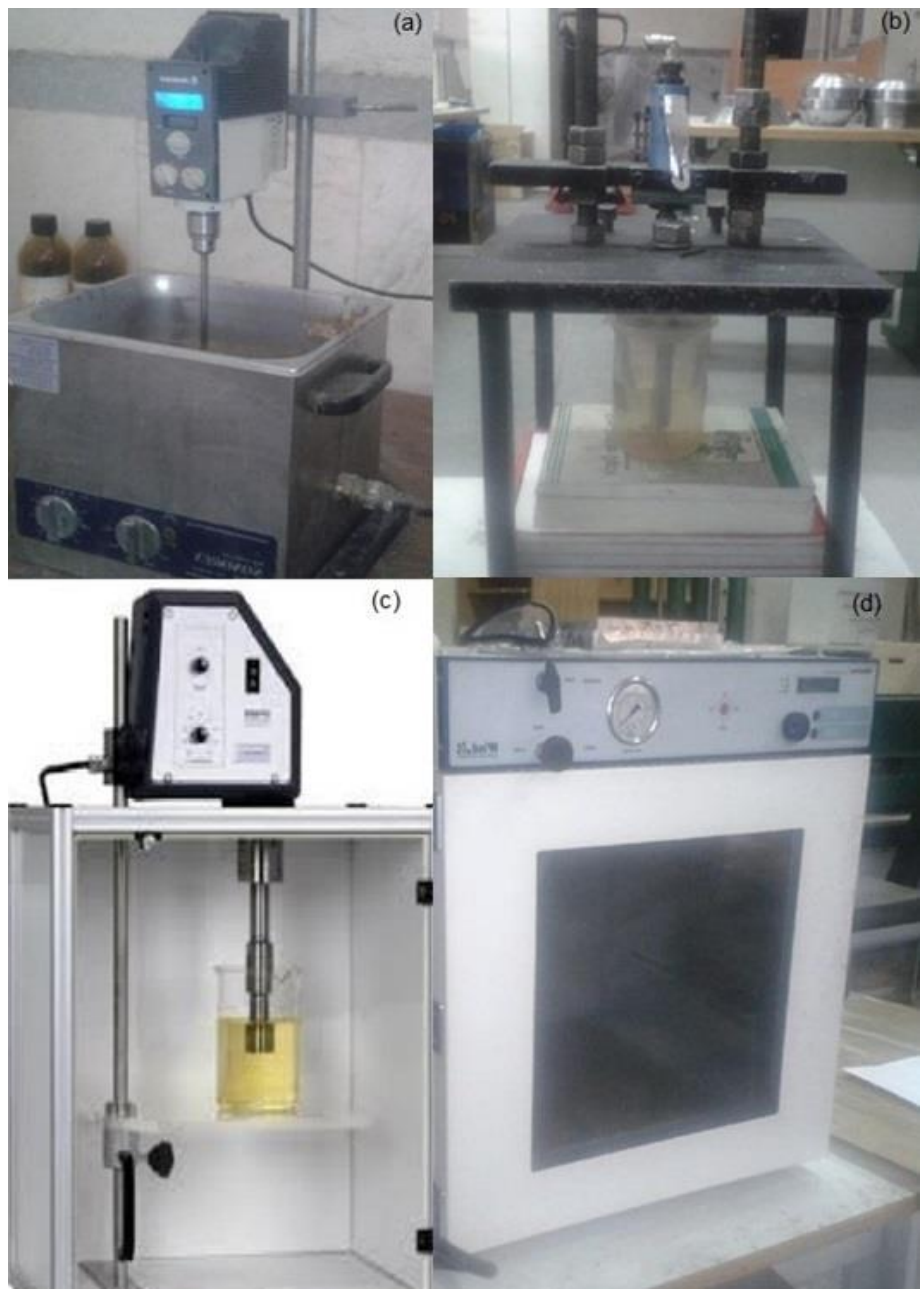

Figure. 1. Four stages of STF preparation a) mechanical mixer, b) high speed shear mixer, c) ultrasonic homogenizer, d) vacuum oven

## 2.5. Impregnation of foams with STF and Glycerol

A fluid injection device has been designed to saturate foams by Newtonian and non-Newtonian fluids because of their extreme viscosity. The fluids were impregnated into the foams when a slow suction of the fluids generated by creating the pressure gradient across the end of the samples immersed inside the specific fluid. The volume fraction of the fluid inside of the foam was determined by weight measurement at the end of the process. Fig. 2 shows the fluid injection instrument. By placing the foams into the device, air pressure at the top of the foams have been increased, but at the bottom surface of the foams decreased by installing a vacuum pump. With this pressure gradient, fluids can be infused through the foams and they are soaked and saturated.

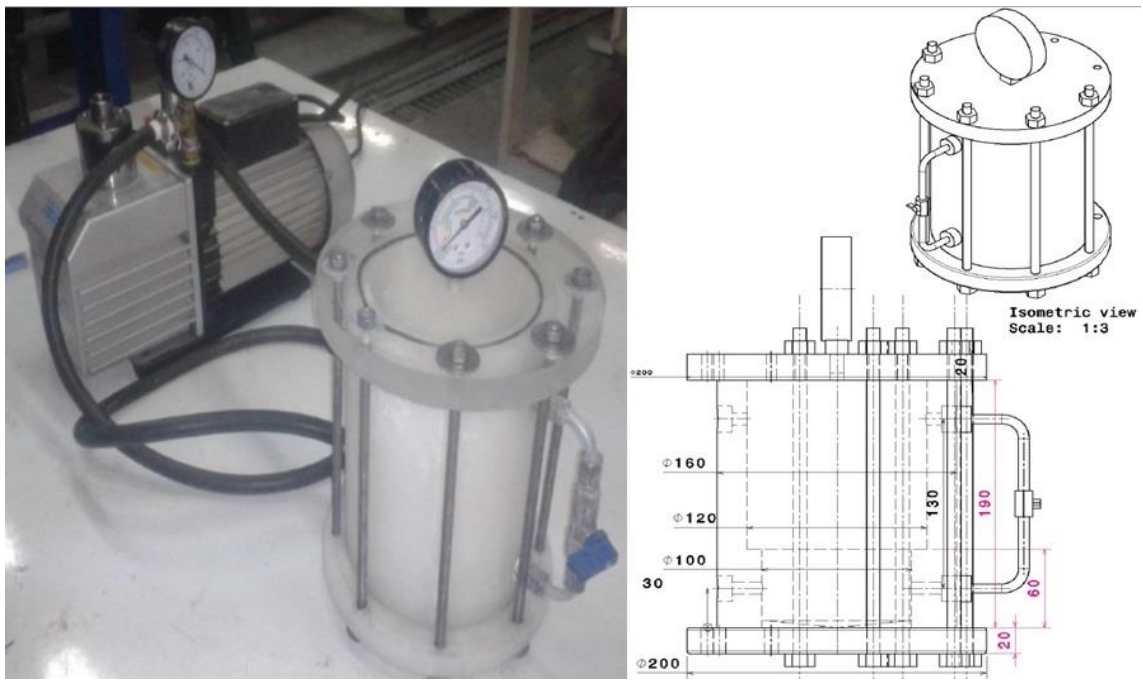

Figure. 2. Schematic and manufactured STF injection device

## 2.6. Quasi static punch shear test (QS-PST)

Quasi-static punch shear test has been carried out using universal Instron testing machine under displacement control at a constant rate of (0.02 mm/sec). The indenter was a surface hardened spherical-nosed cylindrical steel with a diameter of 10 mm and a shank length of 10 cm. Simply-supported and clamped (fully-fixed) boundary conditions were applied to the specimens. The force and displacement of the indenter throughout the test were recorded using a computer connected to the quasi-static machine. The tests were continued until complete penetration of indenter occurred. Figs. 3 and 4 indicate a typical specimen and two different boundary conditions and schematic of the quasi-static punch shear test devices, respectively. Table 1 represents the different testing set up, boundary conditions, and other characteristics of the specimens with their designations.

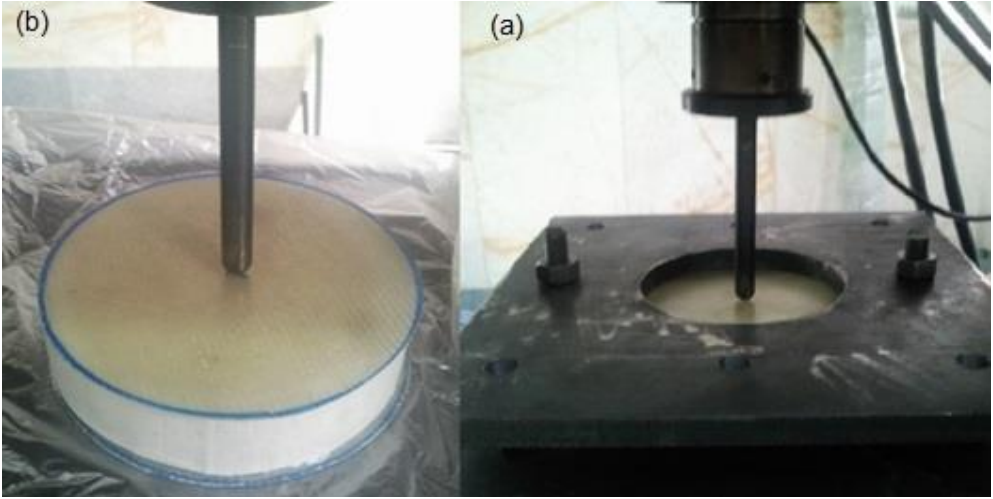

Figure. 3. Quasi-static punch shear test set up and two different boundary conditions, a) clamped (fully-fixed), b) simply supported

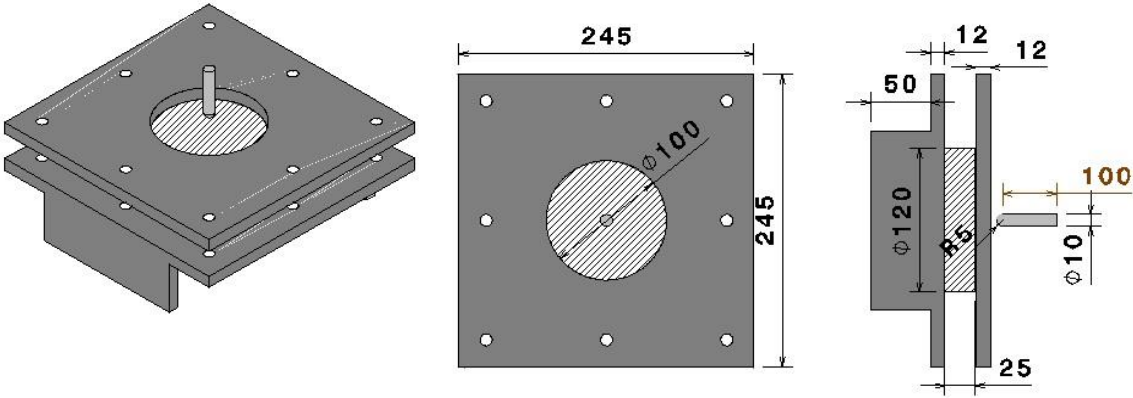

Figure. 4. Schematics of the quasi-static punch shear test with clamped fixture (all dimentions are in mm)

TABLE 1. Characteristics of the specimens and their designated name.

| Name | Number of layer in upper face | Number of layer in Lower face | Core type | Boundary Condition |
|---|---|---|---|---|
| **Emp88ss** | 8 | 8 | Unfilled foam | Simply-Suppo`rted |
| **Emp88ff** | 8 | 8 | Unfilled foam | Clamped (fully-fixed) |
| **Emp44ss** | 4 | 4 | Unfilled foam | Simply-Supported |
| **Emp44ff** | 4 | 4 | Unfilled foam | Clamped (fully-fixed) |
| **Gly88ss** | 8 | 8 | Filled with Glycerol | Simply-Supported |
| **Gly88ff** | 8 | 8 | Filled with Glycerol | Clamped (fully-fixed) |
| **STF88ss** | 8 | 8 | Filled with STF 27.5 wt.% | Simply-Supported |
| **STF88ff** | 8 | 8 | Filled with STF 27.5 wt.% | Clamped (fully-fixed) |

## 3. Results and discussion

### 3.1. Rheological properties

Fig. 5 represents the shear viscosity and shear stress as a function of shear rate for different contents of nanosilica (20, 27.5, and 35% wt.%) in the ethylene glycol base. Literatures investigated that STF with 18% or higher silica content in EG, indicated shear thinning behavior at low shear rate, and extreme shear thickening characteristics at high shear rates (above 100-1) [38].

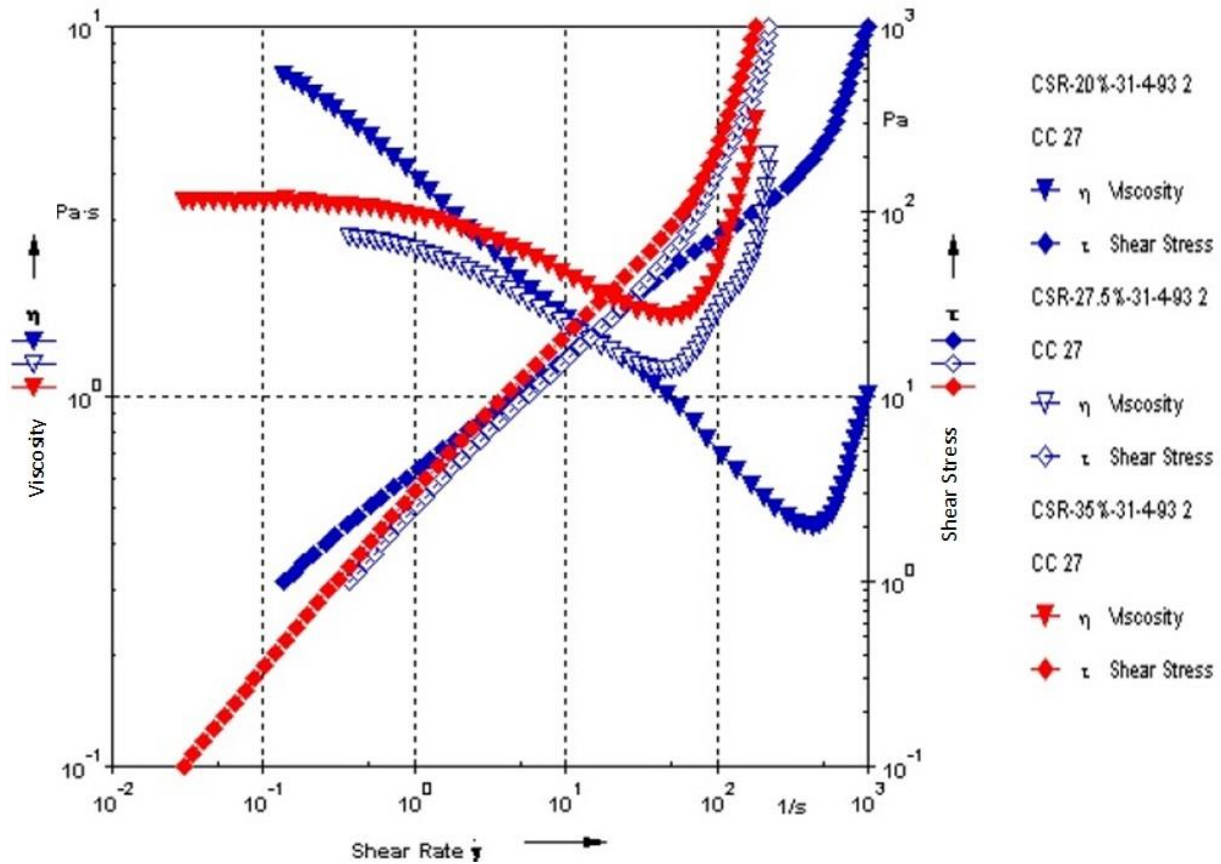

Figure. 5. Viscosity and shear stress as a function of shear rate with three fumed silica concentration

This graph can be divided into three parts. At low shear rates, viscosity of the suspension was not sensible to the shear rate, and fluid acted like a Newtonian fluid; on the other hand, shear thinning behavior has been observed. At low shear rates, nanosilica particles filled the empty spaces between ethylene glycol and some accidental collisions happen between them. With increasing the shear rate, viscosity decreases. In this area, nanoparticles flowed in a certain and specific path. Shear thickening behavior have taken place with the increasing of the shear rate above critical shear rate. The critical shear rate reduced, with the raising of the silica particle concentration. On the other hand, the onset of shear thickening transmitted to lower shear rate when the suspension concentration increased [48]. The critical shear rates are 800 s-1, 80 s-1, and 70 s-1 for the STFs having silica content of 20, 27.5, and 35 wt%, respectively. This critical transition from the flowing liquid to a solid-liquid material is due to the foundation and propagation of shear induced transient aggregates, or hydrocluster, that dramatically increase the viscosity of the fluid [49]. These hydroclusters create from microstructural change where hydrodynamic force overcome inter-particle forces [50]. The onset of hydrocluster caused the transient fluctuation of particles, which followed by the friction between particles and propels to increase in fluid viscosity [51]. In this study, the 27.5 wt% concentration was selected for the STF used to saturate ceramic foams, due to its moderate viscosity. On the other hand, besides the presence of shear thickening behavior at this concentration, fabrication and impregnation of the ceramic foams is easier compared with the suspension by 35 wt% concentration.

## 3.2. Quasi-Static Punch Shear Test

Force-displacement curve of the empty sandwich panel ceramic foam core specimens under different boundary conditions are presented in Fig. 6.

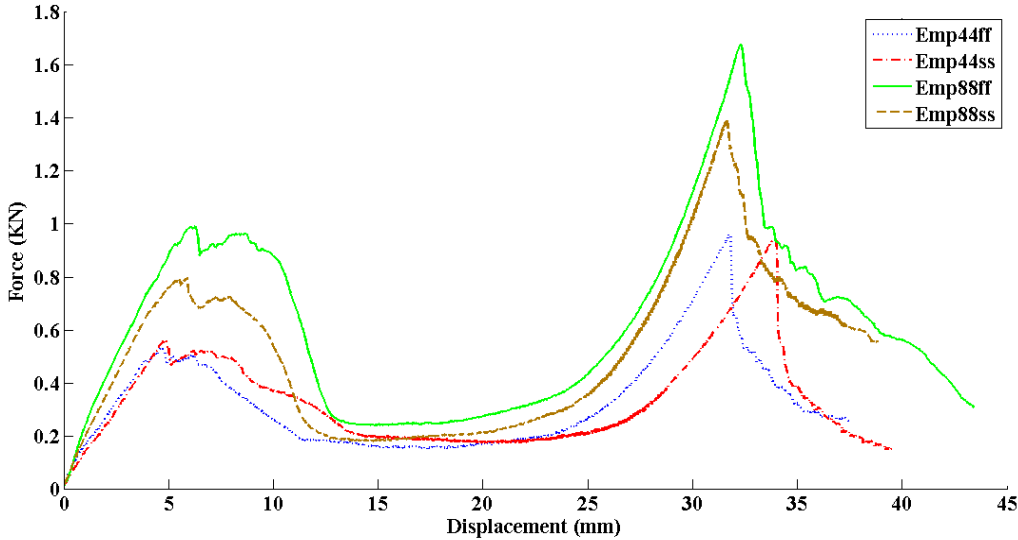

Figure. 6. Force-displacement curve for ceramic foam core sandwich panels

Each of the four curves followed a similar pattern and indicated the following key features: (1) The force increases linearly until reaches to its first lower pick. During the first step, upper layer stretched and bent. (2) Subsequently, enhancing the force in the first step ensued by decreasing its amount because of the crack initiation and propagation in the upper layer. (3) In the third steps, bending of the lower layer increased the force to its second pinnacle at a higher pick. As the indenter continued to perforate, small cracks began to initiate in the lower skin; As a results, force decreases and the only force is the friction between the indenter shank and sandwich panel.

Fig. 7 shows the quasi-static punch shear tests' results for the ceramic foam core sandwich panels saturated with glycerol and shear thickening fluid for different boundary conditions. Glycerol filled foam core sandwich panels behave similar to the unfilled sandwich panels, but with the lower recorded load. One of the main reasons for this reduction is the decreasing of the shear resistance of the foam core due to presence of glycerol which moisten the foam. In the STF impregnated foam core sandwich panels resistant force boosts because of the viscosity and adherence characteristics of the fluid. As the shank continued to perforate, conical plugging started to form which interrupted the fluid flow and shear resistance in the core contributes to reduction in the force until bending of the lower layer of the sandwich panel.

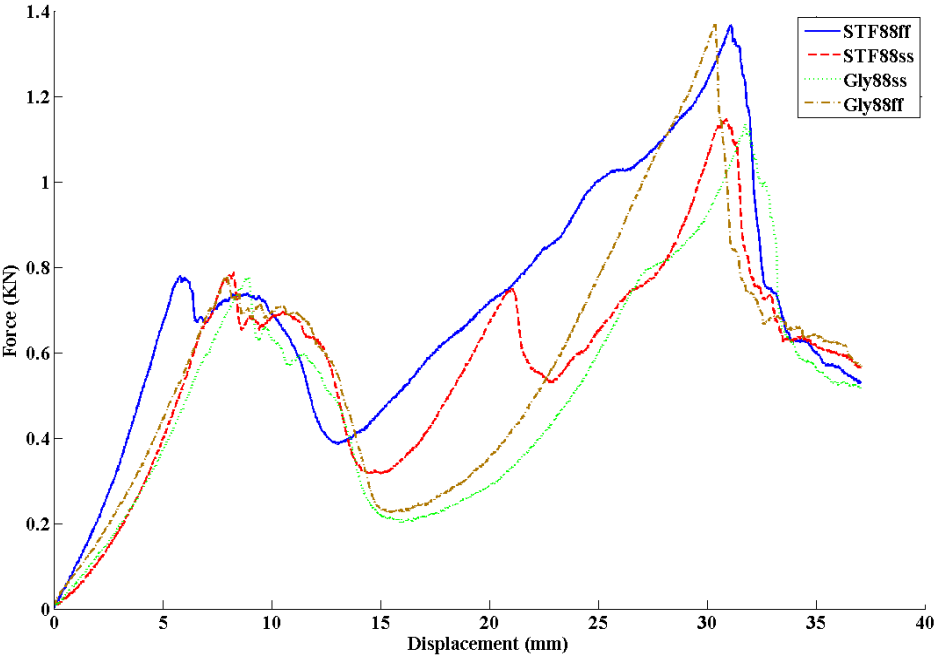

Figure. 7. Force-displacement behavior for ceramic foam core sandwich panels filled with Glycerol and STF during shank penetration in QS-PST

### 3.3. Energy absorption in quasi-static punch shear test

The area under force-displacement curve demonstrates the energy absorption during the penetration process. This could be represented by the following equation.

$$W = E_D = \int_0^{X_f} F(x)dx \tag{1}$$

Variations of energy absorption in the foam core sandwich panels, with or without fillers, are shown in Figs. 8 and 9. It is obvious in Fig. 8 that unfilled foam core sandwich panels behave similarly. In addition, with the increase of supporting restraints, energy absorption increased. Composite layers play a crucial role in energy absorption in sandwich panels during the indentation; moreover, supporting conditions affect the curvature and potential energy of the plate. As a result, with the decreasing of the composite layers from 8 to 4, potential energy reduced, so the influence of supporting condition is negligible. Therefore, supporting conditions have not any tangible effect on 4 layers composite plates, and could be overlooked.

Energy absorption pattern in the 27.5 wt.% STF filled sandwich panels is analogous to the unfilled foam core panels as clear in Fig. 9. STF with further support conditions (clamped) absorbed higher energy during perforation. Energy absorption in this kind of sandwich panels is much higher compared with other sandwich panels during the perforation of the upper layers.

Energy absorption in the unfilled foam core sandwich panels is equal to the 27.5% of treated foam core. Foam at the bottom of the composite plates became tough and brittle with the presence of STF, and composite layers absorbed lower energy due to little deformation resulted from toughening behavior of the foam, but when the indenter passed from the core, higher energy absorbency taken place due to fluid flow.

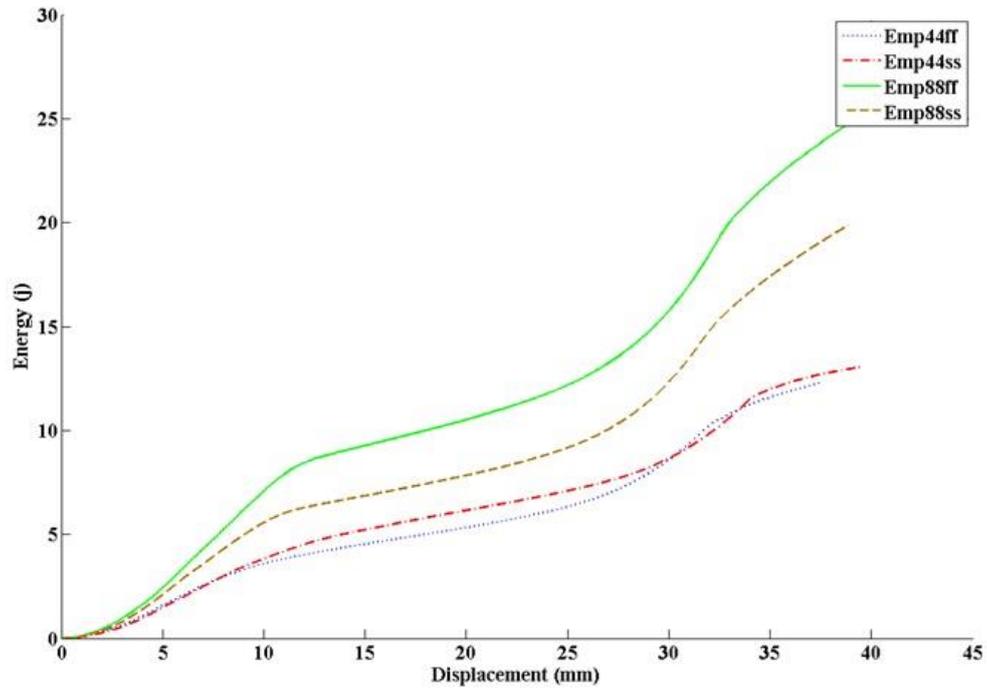

Figure. 8. Energy absorption in the unfilled foam core sandwich panels in QS-PST

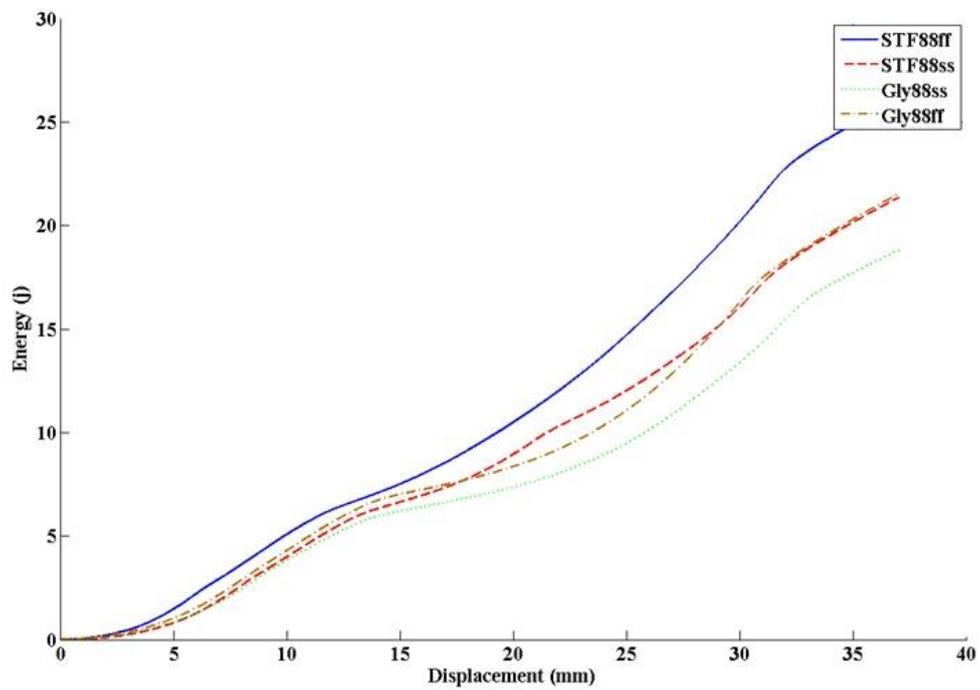

Figure. 9. Energy absorption in the 27.5% treated foam core sandwich panels in QS-PST

### 3.4. Deformation of sandwich panel

*Deformation of top face sheet:* Localized deformation under the indenter was observed at the outset of each test. Firstly small cracks was initiated and propagated under the indenter ensued by force decreasing (Figs. 6 and 7). With the complete fracture of top face sheet, other mechanisms such as crack propagation, delamination, and the rotation of the formed petal were observed and these phenomena contributed to energy absorption. For thinner composite skin, force dropped in a small amount, while the force dropped by a larger amount in a thicker composite skin (In this case study the composite face sheets with 8 layers). In a simply supported specimens debonding between top face sheet and foam core was observed during perforation. Fig. 10 shows post-test photograph of top face sheet of the unfilled foam core sandwich panel.

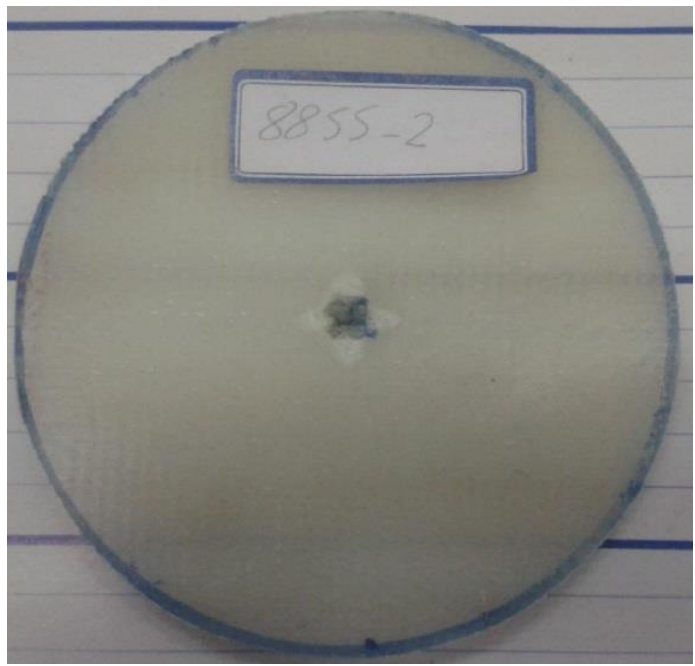

Figure. 10. Top face sheet of the unfilled foam core sandwich panel after quasi-static test

*Deformation of the core:* Opening of the petal in the top face sheet, caused localized damage in the upper face of the foam core. When the indenter reached to the core, it perforated the core in a circular patter with the same radius of the indenter shank. Neat foam core showed a steady increase in displacement (Fig. 6) without increasing in force, but in STF impregnated foam core, force increased with a slight slip during this stage (Fig. 7). Finally, imperfect cone shape was created. Fig. 11 demonstrates upper and lower face of the unfilled foam core.

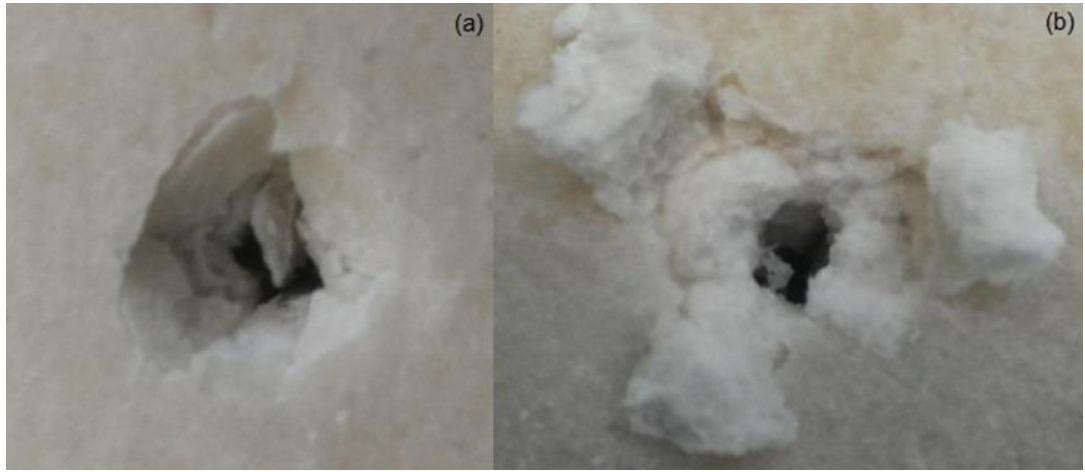

Figure. 11. Quasi-static punch shear test in the unfilled foam core, a) upper face, b) lower face

*Deformation of the bottom face sheet:* Near the second peak (Figs. 6 and 7) cracks were initiated and could be observed in the bottom face sheet. The bottom face-sheet cleaved into three or four pieces under the indenter. Bottom skin of the unfilled foam core sandwich panel is presented in Fig. 12. After complete failure of the bottom face-sheet (passing the indenter through the face-sheet), the force dropped and friction between indenter shank and sandwich panel was the only remaining active force.

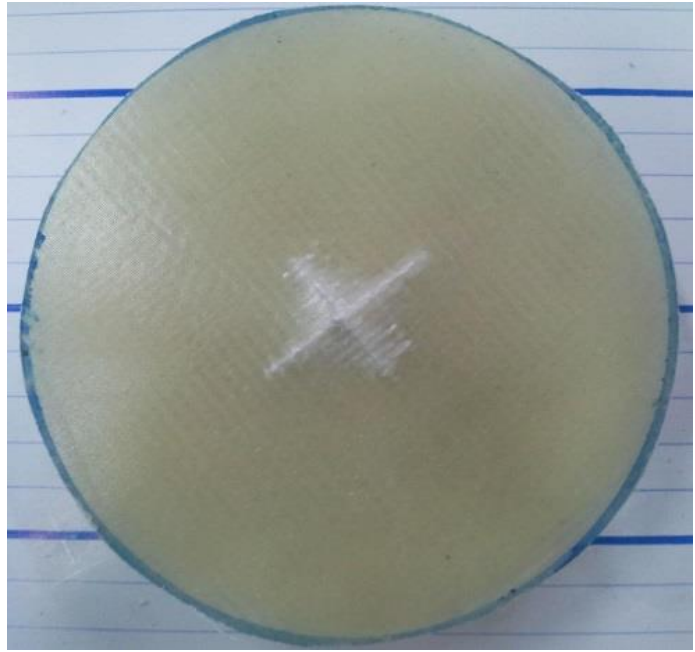

Figure. 12. Bottom face sheet of the unfilled foam core panel after quasi-static test without shank perforation

## 3.5. Damaged area

Fig. 13 compares the damaged area of the top and bottom face-sheets in sandwich panels for unfiiled foam core and filled foam core with glycerol or STF. Also, the effects of boundary condition on the damaged area are visible. Damage area of 8 layers face-sheets of the simply supported sandwich panels filled with STF was the same as unfilled sandwich panels and is about 3% higher than the clamped sandwich panels. Adding glycerol reduced the damaged area significantly; on the other hand, damaged area of the neat foam core and STF saturated foam core is about 119% higher than the damage area of the glycerol foam core.

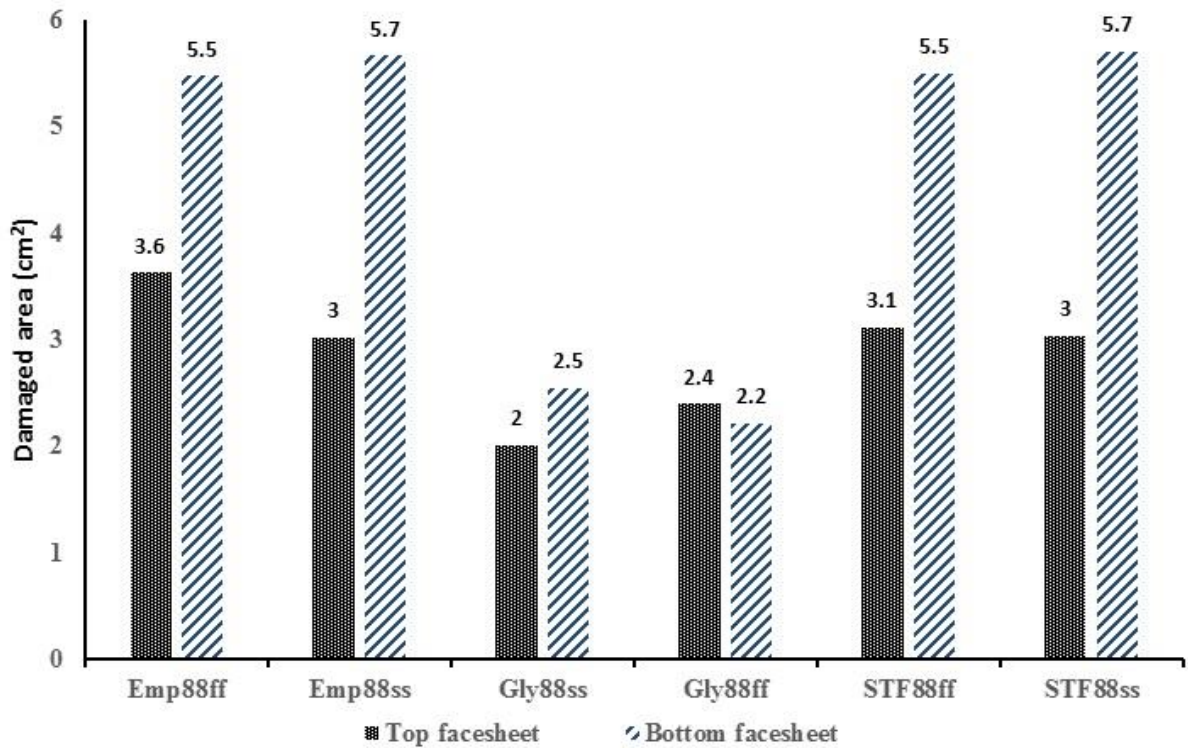

Figure. 13. Comparison of the damaged area in quasi-static punch shear tests for different core foams

Increasing restraint of the structures influenced their stiffness and made them to become more stiffen which contribute to initiating the smaller cracks instead of bigger ones leads to reduction in damaged area. Also, foam under the top face-sheet increase the composites' stiffness, so it limited the crack propagation, led to localized plugging and rotation of the edge of the plug. Moreover, this phenomenon caused composites' matrix were collapsed and damage area increased. Same results could be explained for the glycerol filled foam core.

Fig. 14 indicates the effect of boundary condition and thickness of the face-sheets on damaged area in quasi-static test. This graph can be analyzed from two aspects. 1) For clamped (fully fixed) boundary condition the damaged area of the top face-sheet in 8 layers composite was 25% higher than the 4 layers composite. In addition, this result for the bottom layer was 86%. 2) For simply-supported boundary condition the damaged area of the top face-sheet for 8 layers composite was the same as the 4 layers composite, but in the bottom face-sheet, damaged area of 8 layers composite was 61% higher than the 4 layers. Note that the bottom layers were not supported with foam core like the top layer, so their stiffness is much smaller than the top face. As a result, cracks propagated freely.

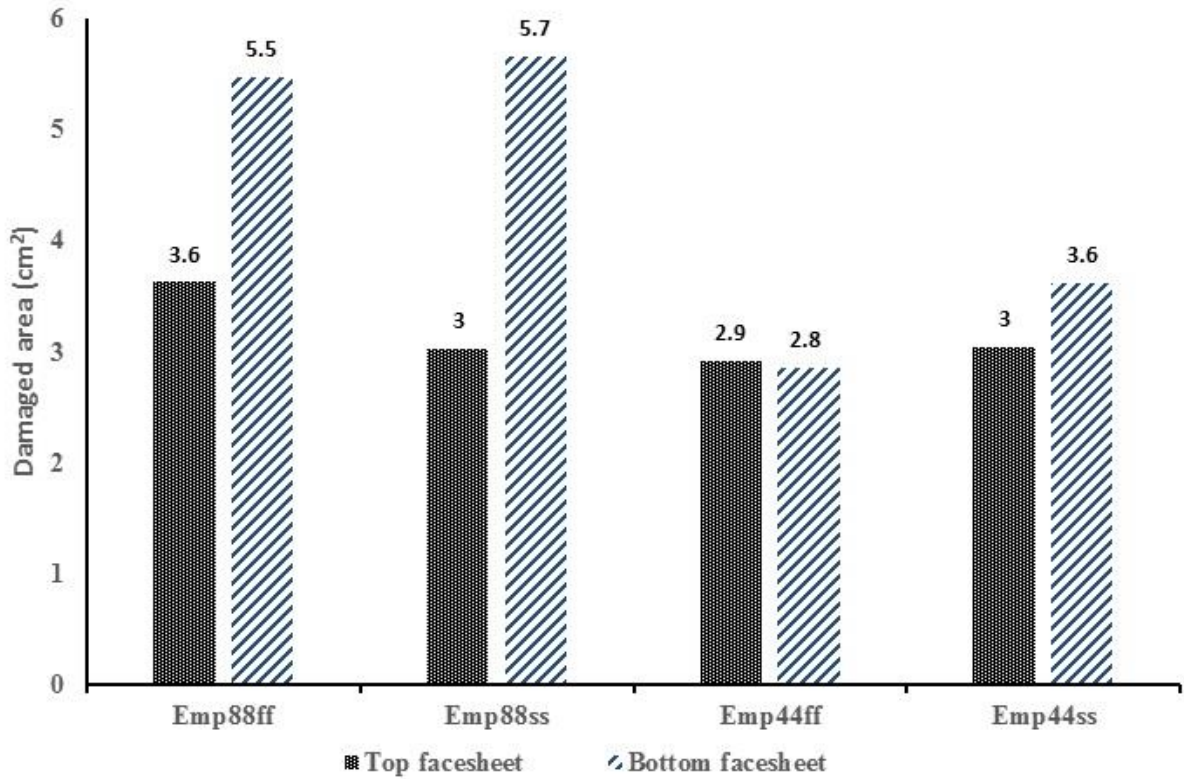

Figure. 14. Effect of boundary condition and thickness of the face-sheets on damaged area in unfilled foam core specimens

## 4. Conclusion

Quasi-static punch shear test (QS-PST) were conducted to study the mechanical response and energy absorption characteristics of the neat foam core sandwich panel, STF filled foam core, and glycerol filled foam core under two different supporting conditions and two different thicknesses of the face-sheets. Rheological tests showed adding foamed silica sharply increased the viscosity and density, but decreased the critical shear rate of the fluid. In the neat foam core sandwich panels, 8 layers face-sheets absorbed the maximum energy and 4 layer face-sheets absorbed minimum energy during indentation. In addition, in 8 layers face-sheet panels, STF filled core had higher indentation resistance Figure. 2. Schematic and manufactured STF injection device compared with Newtonian fluid filled core. Moreover, Damaged area in unfilled foam core sandwich panels was equal to STF saturated foam core when they had simply-supported condition during penetration and Damage area in empty and STF impregnated foam core sandwich panels were much higher than the damage area in glycerol filled foam core panels.

5. **Acknowledgment**

The authors would like to thank Tarbit Modares University for partial financial support of this work.

**References**

[1] Li Z, Zheng Z, Yu J. Low-velocity perforation behavior of composite sandwich panels with aluminum foam core. Journal of Sandwich Structures and Materials. 2012;15:92-109.

[2] Nasirzadeh R, Sabet AR. Influence of nanoclay reinforced polyurethane foam toward composite sandwich structure behavior under high velocity impact. Journal of Cellular Plastics. 2014;52:253-75.

[3] Wang B, Zhang G, Wang S, Ma L, Wu L. High Velocity Impact Response of Composite Lattice Core Sandwich Structures. Applied Composite Materials. 2013;21:377-89.

[4] Gibson LJ, Ashby MF. Cellular solids: structure and properties: Cambridge university press, 1999.

[5] Abisset E, Daghia F, Sun X, Wisnom M, Hallett S. Interaction of inter-and intralaminar damage in scaled quasi-static indentation tests: Part 1–Experiments. Composite Structures. 2016;136:712-26.

[6] Ahmadi H, Liaghat G, Sabouri H, Bidkhouri E. Investigation on the high velocity impact properties of glass-reinforced fiber metal laminates. Journal of composite materials. 2012:0021998312449883.

[7] Aly NM, Saad MA, Sherazy EH, Kobesy OM, Almetwally AA. Impact properties of woven reinforced sandwich composite panels for automotive applications. Journal of Industrial Textiles. 2012;42:204-18.

[8] Andrew JJ, Arumugam V, Santulli C. Effect of Post-Cure Temperature and Different Reinforcements in Adhesive Bonded Repair for Damaged glass/epoxy Composites Under Multiple Quasi-Static Indentation Loading. Composite Structures. 2015.

[9] Baucom J, Zikry M. Evolution of failure mechanisms in 2D and 3D woven composite systems under quasi-static perforation. Journal of composite materials. 2003;37:1651-74.

[10] Gdoutos E, Daniel I. Indentation failure in composite sandwich structures. Experimental mechanics. 2002;42:426-31.

[11] Gunes R, Arslan K. Development of numerical realistic model for predicting low-velocity impact response of aluminium honeycomb sandwich structures. Journal of Sandwich Structures and Materials. 2015;18:95-112.

[12] Zhang T, Yan Y, Li J, Luo H. Low-velocity impact of honeycomb sandwich composite plates. Journal of Reinforced Plastics and Composites. 2015;35:8-32.

[13] Walley S. An introduction to the properties of silica glass in ballistic applications. Strain. 2013;50:470-500.

[14] Erkendirci ÖF, Haque BZG. Quasi-static penetration resistance behavior of glass fiber reinforced thermoplastic composites. Composites Part B: Engineering. 2012;43:3391-405.

[15] Pearson JD, LaBarbera D, Prabhugoud M, Peters K, Zikry MA. Experimental and Computational Investigation of Low-Impact Velocity and Quasi-Static Failure of PMMA. Experimental mechanics. 2012;53:53-66.

[16] Singh AK, Davidson BD, Eisenberg DP, Czabaj MW, Zehnder AT. Damage characterization of quasi-statically indented composite sandwich structures. Journal of composite materials. 2012;47:1211-29.

[17] Sutherland L, Soares CG. The use of quasi-static testing to obtain the low-velocity impact damage resistance of marine GRP laminates. Composites Part B: Engineering. 2012;43:1459-67.

[18] Yu J, Wang E, Li J, Zheng Z. Static and low-velocity impact behavior of sandwich beams with closed-cell aluminum-foam core in three-point bending. International Journal of Impact Engineering. 2008;35:885-94.

[19] Pandya K, Shaktivesh S, Gowtham H, Inani A, Naik N. Shear Plugging and Frictional Behaviour of Composites and Fabrics Under Quasi-static Loading. Strain. 2015;51:419-26.

[20] Feli S, Khodadadian S, Safari M. A modified new analytical model for low-velocity impact response of circular composite sandwich panels. Journal of Sandwich Structures and Materials. 2016;18:552-78.

[21] Roach AM, Evans KE, Jones N. The penetration energy of sandwich panel elements under static and dynamic loading. Part I. Composite Structures. 1998;42:119-34.

[22] Daniel IM, Abot JL, Schubel PM, Luo JJ. Response and Damage Tolerance of Composite Sandwich Structures under Low Velocity Impact. Experimental mechanics. 2011;52:37-47.

[23] Hassan MZ, Cantwell WJ. The influence of core properties on the perforation resistance of sandwich structures – An experimental study. Composites Part B: Engineering. 2012;43:3231-8.

[24] Malcom AJ, Aronson MT, Deshpande VS, Wadley HNG. Compressive response of glass fiber composite sandwich structures. Composites Part A: Applied Science and Manufacturing. 2013;54:88-97.

[25] Studzinski R, Pozorski Z. Experimental and numerical analysis of sandwich panels with hybrid core. Journal of Sandwich Structures and Materials. 2016.

[26] Zhou J, Hassan MZ, Guan Z, Cantwell WJ. The low velocity impact response of foam-based sandwich panels. Composites Science and Technology. 2012;72:1781-90.

[27] Li Z, Zheng Z, Yu J, Yang J. Indentation of composite sandwich panels with aluminum foam core: An experimental parametric study. Journal of Reinforced Plastics and Composites. 2014;33:1671-81.

[28] Jing L, Wang Z, Zhao L. An approximate theoretical analysis for clamped cylindrical sandwich shells with metallic foam cores subjected to impulsive loading. Composites Part B: Engineering. 2014;60:150-7.

[29] Jing L, Wang Z, Ning J, Zhao L. The dynamic response of sandwich beams with open-cell metal foam cores. Composites Part B: Engineering. 2011;42:1-10.

[30] Zhou J, Guan ZW, Cantwell WJ. The impact response of graded foam sandwich structures. Composite Structures. 2013;97:370-7.

[31] Feng D, Aymerich F. Damage prediction in composite sandwich panels subjected to low-velocity impact. Composites Part A: Applied Science and Manufacturing. 2013;52:12-22.

[32] Ruan D, Lu G, Wong YC. Quasi-static indentation tests on aluminium foam sandwich panels. Composite Structures. 2010;92:2039-46.

[33] Schubel PM, Luo J-J, Daniel IM. Low velocity impact behavior of composite sandwich panels. Composites Part A: Applied Science and Manufacturing. 2005;36:1389-96.

[34] Hassan TA, Rangari VK, Jeelani S. Synthesis, processing and characterization of shear thickening fluid (STF) impregnated fabric composites. Materials Science and Engineering: A. 2010;527:2892-9.

[35] Baharvandi HR, Khaksari P, Alebouyeh M, Alizadeh M, Khojasteh J, Kordani N. Investigating the quasi-static puncture resistance of p-aramid nanocomposite impregnated with the shear thickening fluid. Journal of Reinforced Plastics and Composites. 2014;33:2064-72.

[36] Gong X, Xu Y, Zhu W, Xuan S, Jiang W, Jiang W. Study of the knife stab and puncture-resistant performance for shear thickening fluid enhanced fabric. Journal of composite materials. 2013;48:641-57.

[37] Haro EE, Szpunar JA, Odeshi AG. Ballistic impact response of laminated hybrid materials made of 5086-H32 aluminum alloy, epoxy and Kevlar® fabrics impregnated with shear thickening fluid. Composites Part A: Applied Science and Manufacturing. 2016;87:54-65.

[38] Kang TJ, Hong KH, Yoo MR. Preparation and properties of fumed silica/Kevlar composite fabrics for application of stab resistant material. Fibers and Polymers. 2010;11:719-24.

[39] Lomakin E, Mossakovsky P, Bragov A, Lomunov A, Konstantinov AY, Kolotnikov M, et al. Investigation of impact resistance of multilayered woven composite barrier impregnated with the shear thickening fluid. Archive of applied mechanics. 2011;81:2007-20.

[40] Hasanzadeh M, Mottaghitalab V, Babaei H, Rezaei M. The influence of carbon nanotubes on quasi-static puncture resistance and yarn pull-out behavior of shear-thickening fluids (STFs) impregnated woven fabrics. Composites Part A: Applied Science and Manufacturing. 2016;88:263-71.

[41] Fahool M, Sabet AR. Parametric study of energy absorption mechanism in Twaron fabric impregnated with a shear thickening fluid. International Journal of Impact Engineering. 2016;90:61-71.

[42] Park Y, Kim Y, Baluch AH, Kim C-G. Numerical simulation and empirical comparison of the high velocity impact of STF impregnated Kevlar fabric using friction effects. Composite Structures. 2015;125:520-9.

[43] Lu Z, Wu L, Gu B, Sun B. Numerical simulation of the impact behaviors of shear thickening fluid impregnated warp-knitted spacer fabric. Composites Part B: Engineering. 2015;69:191-200.

[44] Bettin G. Energy absorption of reticulated foams filled with shear-thickening silica suspensions: Massachusetts Institute of Technology, 2005.

[45] Bettin G. High-rate deformation behavior and applications of fluid filled reticulated foams: Massachusetts Institute of Technology, 2007.

[46] Ramirez JG. Characterization of shear-thickening fluid-filled foam systems for use in energy absorption devices: Massachusetts Institute of Technology, 2004.

[47] Park Y, Kim Y, Baluch AH, Kim C-G. Empirical study of the high velocity impact energy absorption characteristics of shear thickening fluid (STF) impregnated Kevlar fabric. International Journal of Impact Engineering. 2014;72:67-74.

[48] Na W, Ahn H, Han S, Harrison P, Park JK, Jeong E, et al. Shear behavior of a shear thickening fluid-impregnated aramid fabrics at high shear rate. Composites Part B: Engineering. 2016;97:162-75.

[49] Haris A, Lee H, Tay T, Tan V. Shear thickening fluid impregnated ballistic fabric composites for shock wave mitigation. International Journal of Impact Engineering. 2015;80:143-51.

[50] Lee YS, Wetzel ED, Wagner NJ. The ballistic impact characteristics of Kevlar® woven fabrics impregnated with a colloidal shear thickening fluid. Journal of materials science. 2003;38:2825-33.

[51] Xu Q, Majumdar S, Brown E, Jaeger HM. Shear thickening in highly viscous granular suspensions. EPL (Europhysics Letters). 2014;107:68004.